\documentstyle[12pt]{article}

\input epsf

\begin{document}

\begin{titlepage}
\null\vspace{-62pt}

\pagestyle{empty}
\begin{center}

\vspace{1.0truein} {\Large\bf Quantum-classical interactions
                               through the path integral}

\vspace{1in}
{\large Dimitrios Metaxas} \\
\vskip .4in
{\it Department of Physics,\\
National Technical University of Athens,\\
Zografou Campus, 15780 Athens, Greece\\
metaxas@central.ntua.gr}\\

\vspace{0.5in}

\vspace{.5in}
\centerline{\bf Abstract}

\baselineskip 18pt
\end{center}

I consider the case of two interacting scalar fields, $\phi$ and
$\psi$, and use the path integral formalism in order to treat the
first classically and the second quantum-mechanically. I derive the
Feynman rules and the resulting equation of motion for the classical
field which should be an improvement of the usual semi-classical
procedure. As an application I use this method in order to enforce
Gauss's law as a classical equation in a non-abelian gauge theory. I
argue that the theory is renormalizable and equivalent to the usual
Yang-Mills as far as the gauge field terms are concerned. There are
additional terms in the effective action that depend on the Lagrange
multiplier field $\lambda$ that is used to enforce the constraint.
These terms and their relation to the confining properties of the
theory are discussed.

\end{titlepage}

\newpage
\pagestyle{plain}
\setcounter{page}{1}
\newpage

\section{Introduction}

Consider the case of two scalar fields, $\phi$ and $\psi$,
interacting with the action
\begin{equation}
S(\phi,\psi)=\int d^4 x \,  L(\phi, \psi)
\end{equation}
where
\begin{equation}
 L =  \frac{1}{2}(\partial\phi)^2
               -\frac{1}{2} m^2 \phi^2
               -\frac{g}{4!}\phi^4
               +\frac{1}{2}(\partial\psi)^2
               -\frac{1}{2} M^2 \psi^2
               -\frac{g'}{2} \phi^2 \psi^2
               -\frac{g''}{4!} \psi^4.
\label{S}
\end{equation}
The classical equation for $\phi$ is obtained from
\begin{equation}
\frac{\delta S}{\delta \phi} = -(\Box + m^2)\phi
                                -\frac{g}{3!}\phi^3
                                - g' \phi\psi^2=0.
\label{phi1}
\end{equation}
If one wishes to treat $\psi$ quantum-mechanically the usual,
semi-classical, procedure leads one to replace all terms containing
$\psi$ in (\ref{phi1}) by their quantum expectation values. There
are, however, several problems associated with this procedure,
conceptual as well as technical: One may use a background field
method in order to derive the first (one-loop) quantum corrections
to the effective potential and associated mass terms. If, however, a
higher order calculation is needed, one would have to modify the
propagator using the one-loop result, and be careful in order to
avoid double-counting for any relevant diagrams. The procedure
depends on the model examined and on relative order-of-magnitude
estimates.
 One would like to have a more
self-consistent approach in order to incorporate quantum effects on
classical fields and vice-versa.

Here I use the path integral formalism in order to describe this
problem and I get the resulting Feynman rules that enable one to
study the interactions between the classical and quantum fields. The
result is what one would expect: the classical field propagates only
in tree diagrams, not in loops, and the quantum field propagates as
usual, providing the quantum corrections. This leads to an effective
equation of motion for the classical field that should be an
improvement upon the semi-classical procedure. The Feynman rules
presented here reorganize the entire perturbation series and enable
one to treat these problems in a self-consistent manner that can
easily be extended in other models.

The path integral approach has been used before in order to treat
pure classical mechanics \cite{gozzi, mauro} and investigate
problems of classical behavior in quantum field theory
\cite{mueller, jon}. The formalism developed here has many
similarities to these previous works. The main method is an
extension of \cite{gozzi, mauro} to the case of interactions between
classical and quantum fields (there are, however, some important
differences even in the case of purely classical fields).

In Sec.~2 I develop the main formalism, derive the Feynman rules and
get an effective action from which the equation of motion for the
classical field can be obtained. An important advantage of this
method is that it is a self-consistent procedure that can be used in
order to calculate higher order effects.

 In Sec.~3 I use this method in
order to treat Gauss's law as a classical equation in a non-abelian
gauge theory. Because of the asymmetry of the Feynman rules
described here, the effective action contains, besides the usual
Yang-Mills terms, additional terms that depend on the Lagrange
multiplier field that is used to enforce the classical constraint.
These terms are of the Coleman-Weinberg type \cite{CW}, and are
reminiscent of phenomenological effective actions that have been
used in order to describe confinement \cite{kogut, bander, thooft}.
They have similar interpretation here, where one can also see the
region of validity of perturbation theory.

Since the method presented here is new I include two Sections of
comments. In Sec.~4 I discuss the applications to non-abelian gauge
theory and in Sec.~5 I make some general comments regarding this
method.

\section{Path integral and effective action}

In order to calculate
\begin{equation}
Z(J,J')=\int[d\phi][d\psi]\delta(\phi-\phi_{\rm cl}) \exp\left(
i\int L(\phi, \psi) + J\phi +J' \psi \right),
\end{equation}
where $\phi_{\rm cl}$ is the solution of (\ref{phi1}), I use the
Lagrange multiplier $\lambda$ and ghost fields $c$, $\bar{c}$,
similarly to the work in \cite{gozzi, mauro}, and adding another
source $\Lambda$, the path integral to be evaluated becomes
\begin{equation}
Z(J, J', \Lambda) = \int[d\phi][d\psi][d\lambda][dc][d\bar{c}]
                      \exp{(i\int \tilde{L}+J\phi+J'\psi+\Lambda\lambda)}
\end{equation}
where
\begin{equation}
     \tilde{L}=L+ \lambda \frac{\delta
     S}{\delta\phi} + \bar{c}\frac{\delta^2 S}{\delta\phi^2}c
\label{Stilde}
\end{equation}
is the modified Lagrangian with the corresponding modified action
$\tilde{S}$.
 For the simplest case of two interacting scalar fields with
(\ref{S}) we get
\begin{equation}
\tilde{S}=S+\int_x \left( \lambda K \phi +\bar{c} K c -
          \frac{g}{3!} \lambda \phi^3 -g'\lambda\phi\psi^2 +
         \frac{g}{2} \bar{c} \phi^2 c + g'\bar{c}\psi^2 c
         \right),
\label{Sphi}
\end{equation}
where $K = -(\Box + m^2)$. The propagators and the vertices can be
deduced from here. For the $\lambda$ and $\phi$ fields we get
\begin{equation}
\int[d \phi][d\lambda]e^{ i \int \frac{1}{2}\phi K \phi + \lambda K
\phi
                            + J\phi +\Lambda\lambda} = N
    e^{-\frac{i}{2}\int(2JG\Lambda -\Lambda G \Lambda)},
\label{basic}
\end{equation}
where $N$ is a normalization factor, independent of the sources, and
 $G=1/(k^2-m^2 +i\epsilon)$, in momentum space, is the usual Feynman propagator.
Accordingly there is no $\phi - \phi$ propagator, there are,
however, a mixed $\lambda - \phi$ propagator equal to $G$, and a
$\lambda - \lambda$ propagator equal to $-G$. The remaining $\psi -
\psi$ and ghost propagators as well as the various  vertices are as
usual from (\ref{Sphi}).

One can now check: at one loop order the loops with the
$\lambda-\phi$ propagator cancel with the ghost loops and similar
cancelations exist in higher loops, loops with the $\lambda-\lambda$
propagator do not appear because of the Feynman rules of the
modified action (the vertices are at most linear in $\lambda$)
 with the final result that the $\phi$
field does not have quantum corrections but only propagates
classically through tree diagrams. A typical line with the $\lambda$
and $\phi$ fields is either the sum of two $\lambda-\phi$ and one
$\lambda-\lambda$ propagator, or a single $\lambda-\phi$ propagator,
in both cases equal to $G$. The $\psi$ field, of course, propagates
also in loops as a genuine quantum field  and gives the quantum
corrections to the classical field $\phi$.

I should note here that the propagator $G$ that we get for the
classical field is the Feynman propagator, and not the retarded one
that is usually employed in classical mechanics. There are two
reasons for that: first, the boundary conditions used in the path
integral (the field goes to zero at infinity) are different than the
ones usually employed in classical mechanics (the field
configuration is given at an initial time). One can check with a
more careful evaluation of (\ref{basic}) that we get, indeed, the
Feynman prescription. This is true even if we have only the
classical field in our theory. One can presumably use the path
integral formalism with different boundary conditions in order to
attack purely classical problems. Then the retarded propagator would
probably emerge, as in \cite{mauro}. A second, physical reason that
is relevant here, is that we want to study interactions between the
classical and quantum fields. The possibility of particle creation
and annihilation is essential for both the classical and quantum
fields. Our classical field, therefore, admits both particles and
antiparticles propagating classically.

Renormalization of the theory proceeds as usual. All divergent terms
come from loops of the quantum field $\psi$ and depend on the
quantum-classical coupling $g'$ or the quantum coupling $g''$, not
on $g$. The classical parameters $g$ and $m$ get also renormalized.
However, pairs of terms like $g\lambda\phi^3$ and $g\phi^4$ in the
modified action $\tilde{S}$ have the same divergencies associated
with, accordingly the renormalization procedure does not affect the
classical nature of the field $\phi$.

I will now proceed to show how the classical equation for $\phi$
gets modified in the presence of quantum interactions. If we use the
generating functional $Z(J, J', \Lambda)$ to construct $W(J, J',
\Lambda)$, the generating functional for connected diagrams, and
from that $\Gamma(\phi, \psi, \lambda)$, the effective action, we
see that the appropriate equation is
\begin{equation}
\left(\frac{\delta\Gamma}{\delta\lambda}\right)_{\lambda=0}=0.
 \label{qc}
\end{equation}
This can be verified if we consider the theory with only the
classical field and use the properties of the Legendre
transformation in order to express the classical equation $\delta
S/\delta\phi = -\Lambda =0$ in terms of the effective action. So the
equation of motion for the classical field $\phi$ can be taken from
the tadpole one-particle irreducible graphs with one external field
$\lambda$ that contain $\psi$ loops and lines of the $\lambda$ and
$\phi$ fields, but not loops of the classical field. For the case of
the two interacting scalar fields of (\ref{S}) we get at one loop
order
\begin{equation}
-(\Box + m^2)\phi =\frac{\partial V_{\rm eff}(\phi, \psi)}{\partial
\phi} \label{result1}
\end{equation}
where
\begin{equation}
V_{\rm eff} = \frac{g}{4!}\phi^4 +  \frac{g'}{2}\phi^2 \psi^2
  + \frac{i}{2}
          \int\frac{d^4 k}{(2\pi)^4}\ln\left(\frac{(k^2-m_{\phi}^2)(k^2-M_{\psi}^2)+(g'\phi\psi)^2}
             {(k^2-m_{\phi}^2)} \right)
\label{result2}
\end{equation}
is the usual one loop effective potential of the original theory
without the $\phi$ loop ($m_{\phi}^2 = m^2 +\frac{1}{2}g\phi^2
+g'\psi^2$, $M_{\psi}^2 = M^2 + g'\phi^2 + \frac{1}{2}g''\psi^2$).
In fact (\ref{result2}) also contains the loop with only the $\psi$
field. Since this is, however, $\phi$-independent it does not
contribute to (\ref{result1}). One can check that when the
quantum-classical coupling $g'$ is zero this reduces to the ordinary
equation for a classical Klein-Gordon field. Regularization and
renormalization are performed as usual, as was discussed above.

This result can, of course, be also obtained using a background
field perturbation theory. However, if one wishes to study higher
order corrections, perturbation theory has to be reorganized at any
order so as to incorporate the previous order result and avoid any
multiple-counting. This problem does not appear here since we have a
well-defined effective action, that depends on the auxiliary field
$\lambda$, with fixed Feynman rules from the beginning.

The resulting equation for the classical field is written as an
equation of motion from an effective action that contains what one
would expect: arbitrary loops of the quantum field but not loops of
the classical field. It depends on the vacuum expectation value of
the quantum field $\psi$, enabling one to also study problems of
symmetry breaking. Even for $\psi=0$, of course, it provides the
quantum corrections to the classical equations of motion. What is
more important, this method allows one to include higher order
corrections self-consistently through the effective action
formalism.

The complete effective action also describes the quantum properties
of the field $\psi$. One can then determine the effects of the
classical field on the quantum field via its effective potential or
other terms. There are also higher order terms that involve powers
or derivatives of the auxiliary field $\lambda$. Their relevance, if
any, to the combined dynamics of the system is not clear from this
work. In the simple example described here, since there is no
symmetry breaking, we have to set $\lambda =0$ anyway in order to
derive the effective equation of motion. In cases with symmetry
breaking, however, these terms may turn out to be important. One
interesting case will be described in the next Section in the
context of the non-abelian gauge theory, where they may be relevant
to the confining properties of the theory.

\section{An application in non-abelian gauge theory}

The formalism of the preceding section applies strictly in the case
of two interacting fields. Here, however, I will show how it can be
used in the case of the non-abelian gauge theory, in order to treat
Gauss's law as a classical equation. The strategy will be the
following: I will choose first a non-covariant gauge in which pure
Yang-Mills is well-defined and renormalizable, and use the Feynman
rules derived here in order to map a sector of the theory onto the
usual Yang-Mills, and identify the remaining (or missing) terms as
additional contributions to the effective action, that depend on
$\lambda$.

For the non-abelian gauge theory with action $S$ and Lagrangian
\begin{equation}
 L=-\frac{1}{4}F^a_{\mu\nu}F^{a\mu\nu}
 \label{YM1}
\end{equation}
where $F^a_{\mu\nu}=
\partial_{\mu}A^a_{\nu}-\partial_{\nu}A^a_{\mu}+gf^{abc}A^b_{\mu}A^c_{\nu}$,
I will use Lagrange multipliers $\lambda^a$ and ghost fields
$\bar{c}^a, c^a$, and get the modified action
\begin{equation}
\tilde{S} = S + \int \lambda^a \frac{\delta S}{\delta A_0^a} +
\bar{c}^a\frac{\delta^2 S}{\delta A_0^a \delta A^b_0}c^b
 \label{YM2}
\end{equation}
in order to treat Gauss's law
\begin{equation}
\frac{\delta S}{\delta A_0} = - D\cdot\vec{E} =0
 \label{gauss}
\end{equation}
classically ($S$ and $L$ in this Section will denote the Yang-Mills
values). In the modified action we add a source term $\Lambda^a
\lambda^a$ together with the source terms $J^a_{\mu} A^{a \mu}$, as
before, in order to derive the Feynman rules. We also have to add a
gauge fixing term, so I will choose a non-covariant gauge that is
well-defined and does not depend on $A_0$, namely the axial gauge
fixing term
\begin{equation}
L_{\rm ax}= -\frac{(n\cdot A)^2}{2\xi}
 \label{axial}
\end{equation}
with a purely spatial four-vector $n^{\mu}=(0,\vec{n})$.

The usual Feynman rules for Yang-Mills with action $S$ in the axial
gauge involve the gauge field propagator
\begin{equation}
G_{\mu\nu}^{ab} =\frac{-\delta^{ab}}{k^2} \left( g_{\mu\nu}-
                      \frac{k_{\mu}n_{\nu}+k_{\nu}n_{\mu}}{k\cdot n}
                      +k_{\mu}k_{\nu}\frac{n^2 +\xi k^2}{(k\cdot
                      n)^2} \right)
\label{usual}
\end{equation}
and the usual QCD vertices \cite{lieb1}. The new Feynman rules that
we get for the modified action, using (\ref{basic}), involve an
$A_0-A_0$ propagator
\begin{equation}
\tilde{G}_{00}= G_{00} - G_c
\end{equation}
(I will not denote the color indices where obvious)
 and $A_0-\lambda$ as well as $\lambda-\lambda$ propagators
\begin{equation}
G_{0\lambda} = -G_{\lambda\lambda}=G_c
\end{equation}
where $G_c=1/\vec{k}^2$ is the Coulomb propagator. The remaining
propagators $G_{0i}$ and $G_{ij}$ are the same as the usual
Yang-Mills. That is, the effect of the classical constraint of
Gauss's law has been to split the Coulomb interaction from the
propagator and treat it classically. We also have the usual vertices
of QCD, and an additional set of vertices: for every QCD vertex that
contains $A_0$ we have a vertex where an $A_0$ leg is replaced by
$\lambda$. There is also the ghost sector which is similar to the
previous Section, and will be described shortly. This ghost sector
is not related to the usual Fadeev-Popov ghosts, which I will assume
that decouple \cite{lieb1}.
 We can now discuss
renormalizability and relation to the usual Yang-Mills:

As far as the gauge field terms $A^2$, $A^3$ and $A^4$ terms are
concerned we get the same terms as usual Yang-Mills: what used to be
a $G_{00}$ propagator can now be formed as either the sum of
$\tilde{G}_{00}$, two $G_{0\lambda}$ and one $G_{\lambda\lambda}$,
or $\tilde{G}_{00}$ and one $G_{0\lambda}$, in both cases equal to
the usual $G_{00}$. This can be verified for every diagram. There is
one difference: The ghost loop cancels an instantaneous Coulomb loop
from every diagram in the usual Yang-Mills that contains a closed
$A_0$ loop. I will assume that regularization can be performed so
that these loops vanish, that is, relations like
\begin{equation}
\int_k \frac{1}{\vec{k}^2(\vec{k}+\vec{p})^2} =0
 \label{con1}
\end{equation}
hold, together with usual axial gauge integrals like
\begin{equation}
\int_k \frac{1}{(n\cdot k)^2} =0.
 \label{con2}
\end{equation}
This assumption is supported by results of split dimensional
regularization \cite{lieb2}. In fact, even without this assumption,
it is possible that the theory will be renormalizable, since it is
the same closed loop that is missing from every would-be Yang-Mills
diagram, the verification of this, however, would be highly
non-covariant. In any case, with the previous assumption, the theory
is renormalizable and equivalent to the usual Yang-Mills as far as
the gauge fields are concerned. The divergencies of the remaining
terms $\lambda\, \delta S/\delta A_0$ in the modified action are the
same as their counterparts in $S$, as in the previous Section, so
these are renormalizable too. We now turn to the discussion of the
$\lambda$ terms.

Consider first the one-loop terms with external legs $\lambda$ with
zero momentum, for what would be an effective potential term
$U(\lambda)$. All the propagators can run inside the loop with one
exception: The $A_0-\lambda$ and $\lambda-\lambda$ propagators
cannot appear because the vertices are linear in $\lambda$. An
example of such a diagram is shown in Fig.~1. So the Coulomb
interaction is missing. Had these terms been there we would have the
full covariant propagator $G_{\mu\nu}$ in the loop. All these
diagrams, therefore, would add up to zero (their value would be the
same as an effective potential term for $A_0$ in the usual
Yang-Mills which does not exist because of gauge invariance).
Accordingly the sum of the diagrams that contain a Coulomb
interaction between two external legs $\lambda$ has to be
subtracted. The relevant vertices ($\lambda-A_0-A_i$) are the same
as the QCD vertex ($A_0-A_0-A_i$) and
 the missing terms with a Coulomb interaction running between two
 external legs $\lambda$
correspond to an effective interaction term
\begin{equation}
\delta L = m^2 A_i \frac{k_i k_j}{\vec{k}^2} A_j
 \label{res3}
\end{equation}
(in momentum space, with $k$ the momentum running in the loop) with
\begin{equation}
m^2 = g^2 C \lambda^2
 \label{res4}
\end{equation}
where $f^{acd}f^{bcd} = C \delta^{ab}$, $\lambda^2 =
\lambda^a\lambda^a$. One can add this effective interaction term in
the usual Yang-Mills action in order to derive the terms in the
effective action that depend on $\lambda$. It corresponds to missing
terms so the pieces calculated have to be subtracted. The
calculation is highly non-covariant, however, once we have
identified the effect of the missing diagrams as an effective
interaction term to be added to the usual Yang-Mills action, there
is no reason to continue in the axial gauge in order to complete the
calculation \cite{jackiw}. It is more convenient to choose a Feynman
gauge-fixing term, $(\partial_{\mu}A_{\mu})^2/2$, in which case the
effective propagators become:
\begin{equation}
D_{00}=\frac{-1}{k^2} \label{p1}
\end{equation}
\begin{equation}
D_{ij}=\frac{1}{k^2}
       \left(\delta_{ij}+\frac{m^2 k_i
       k_j}{(k^2-m^2)\, {\vec{k}}^2}\right).
\label{p2}
\end{equation}
The usual Fadeev-Popov ghosts do not contribute and the calculation
of the relevant terms gives the final form of the effective action:
\begin{equation}
\Gamma=\int_x  -Z(\lambda)\,\lambda \,D\cdot\vec{E} -\frac{1}{4}
Z(\lambda)\,F^2+U(\lambda)
 \label{res5}
\end{equation}
with
\begin{equation}
U(\lambda)=c_1\,g^4\lambda^4\left(\ln{\frac{\lambda^2}{\mu^2}}
                                -\frac{1}{2}\right)
 \label{resU}
\end{equation}
\begin{equation}
Z(\lambda)=1+ c_2\, g^2\ln{\frac{\lambda^2}{\mu^2}}
 \label{resZ}
\end{equation}
where $c_1=C^2/64 \pi^2$, $c_2=C/6\pi^2$. We have an effective
action of the Coleman-Weinberg form \cite{CW} with a few
differences:

 First, the potential appears with the opposite sign,
since it corresponds to missing diagrams. The counterterms that were
needed for its renormalization were chosen so that $U''(0)=0$ and
$U'(\mu)=0$. The condition $U'(0)=0$ does not fix the counterterms,
so we have to pick the scale $\mu$ (primes denote derivatives with
respect to $\lambda$).

 The factors
$Z$ in the first two terms of the effective action (\ref{res5}) can
be calculated in the same manner, in terms of missing diagrams. They
 are
expected to be the same by individual diagram inspection. A first
calculation gives the result presented above.
 $Z(\lambda)$ was calculated from the $\lambda$-dependent, $\vec{p}^2$ coefficient, of the
 $A_0-A_0$ wavefunction renormalization diagrams with external
 momentum $p$ (subtracted from the tree level term). Other terms
 ($A_i-A_j$ for example) should give the same value for $Z$. This,
 however, will have to be verified, because of the asymmetry of the
 Feynman rules described here. In any case, for the preliminary
 analysis presented below, the main fact that we need is that
 renormalization conditions can be chosen so that $Z(\mu)=1$.

 Although
there is only one coupling constant in the theory the one-loop terms
are reliable when $g^2\ln(\lambda/\mu)$ is small, since the tree
value for $U$ is $0$ and for $Z$ is $1$.

Another possibility that was not encountered in the simple model of
the previous Section, is the generation of $(\nabla\lambda)^2$
terms, because of the asymmetry of the Feynman rules described here.
The tree level value of this term should be set to zero by
counterterms, it is a possible fine-tuning problem of the method.
Generation of these terms at higher order via the Coleman-Weinberg
mechanism does not affect the analysis presented below.

At $\lambda\approx\mu$ we have $Z\approx 1$ and ordinary
perturbative Yang-Mills. As a first approximation to the effective
action we take the abelian, electric parts, for $Z\approx 1$
\begin{equation}
\Gamma_0 =\int_x \lambda\nabla^2 A_0
                 +\frac{1}{2}A_0\nabla^2 A_0 +U(\lambda).
\label{res6}
\end{equation}
The equations
\begin{equation}
\frac{\delta\Gamma_0}{\delta\lambda} =0
 \label{sem1}
\end{equation}
\begin{equation}
\frac{\delta\Gamma_0}{\delta A_0} =0
 \label{sem2}
\end{equation}
have the simultaneous solution
\begin{equation}
A_0 = -\lambda_B
 \label{sol1}
\end{equation}
with $\lambda_B$ the solution of
\begin{equation}
\nabla^2\lambda = \frac{\partial U}{\partial\lambda}.
 \label{sol2}
\end{equation}
Note, first, that, although similar in form, the two equations
(\ref{sem1}) and (\ref{sem2}) are quite different conceptually: The
first equation is an expression of the classical nature of Gauss's
law and should be satisfied identically to all orders. The second
equation is a usual semiclassical equation, only to be satisfied
approximately.

Equation (\ref{sol2}) has a soliton (bounce, bubble) solution
$\lambda_B(r)$, spherically symmetric in the three-dimensional
radius $r$, similar to the bounce solutions that are associated with
tunneling at zero and finite temperature \cite{coleman, linde},
although here it is not related to either tunneling or finite
temperature. $\lambda_B(r)$ is of order $\mu$ for sufficiently small
$r$, and goes rapidly to zero for $r$ larger than the radius $R_B$
of the bounce (which is of order $\frac{1}{g^2 \mu}$). It
corresponds to a confining potential for the electric field
(\ref{sol1}), reminiscent of a bag model, such that deep inside the
bounce we have ordinary perturbative Yang-Mills with a zero electric
field, and a strong electric field appearing as we get closer to the
bounce radius (in fact, the bounce solution is not of the thin wall
type, it resembles more the three-dimensional thick wall bubbles of
\cite{linde}, and has to be determined numerically).

As we approach the radius of the bubble, however, when $\lambda$
goes to zero, the approximation $Z\approx 1$ does not hold; in fact,
when $\lambda$ is non-perturbatively small $Z$ goes to zero and the
full non-perturbative features of QCD become important \cite{kogut,
bander}. Our perturbative parameter is the same as that for
Coleman-Weinberg models, namely $g^2\ln{(\lambda/\mu)}$, and only
when this is small can the higher loop effects be neglected. It is,
of course, possible to improve on these results with the use of the
renormalization group, and even when this is not the case, the
effective action presented here may be quite useful
phenomenologically.

Another important fact that should be mentioned for the effective
action (\ref{res5}) proposed here, as well as for the "dynamically
induced" mass term (\ref{res4}), is that they are gauge invariant,
provided the auxiliary field $\lambda$ is gauge covariant (under a
gauge transformation $V$ we have $\lambda \rightarrow V \lambda
V^{-1}$). The consequences of this gauge invariance, and its
possible associated BRS symmetries, are not obvious because of the
peculiarities of the Feynman rules described here; this is, however,
another indication in support of the arguments presented above,
namely that the theory is renormalizable and compatible with
ordinary, perturbative Yang-Mills.

\section{Comments}

The upshot of this work, as far as the non-abelian gauge theory is
concerned is the following: there is a sector of the theory, namely
the Coulomb interaction, that is purely classical in nature, by
virtue of Gauss's law; this cannot be expressed self-consistently in
perturbation theory unless one employs a skewed set of Feynman rules
of the type described here. This has the effect of generating
additional terms in the effective action that reveal both the
appearance of the confining properties of the theory and the limits
of validity of perturbation theory.

The appearance of the inverted effective potential term $U(\lambda)$
that was described in the previous Section does not indicate an
energetic instability of the theory, it is, however, related to the
instability of the perturbative vacuum. In fact, there is no direct
Hamiltonian interpretation for the effective action presented here,
$\lambda$ remains an auxiliary Lagrange multiplier that has to be
eliminated via $\delta\Gamma/\delta\lambda =0$. The perturbative
vacuum $\lambda=\mu$, however, cannot exist for all space, since it
has infinite action. A finite action soliton solution that was
presented in the previous Section shows signs of confining behaviour
and, at the same time, gives clues about the region of validity of
perturbation theory.

It is not clear whether this method hints on fundamental results
(problems rather) of the usual quantization procedure of non-abelian
gauge theories, or is purely of a heuristic value, I would like,
however, to make some comments in support of the former: First of
all, something that is obvious, but should be, nevertheless,
mentioned, is that this method does not change at all the rules for
the abelian gauge theory. The Coulomb interaction splits again as
presented here. Since, however, photons do not have any
self-interactions, the result is equivalent to the usual Feynman
rules for abelian gauge theory (the inclusion of fermions is also
straightforward). The usual Fadeev-Popov procedure (abelian or
non-abelian) also is not related and does not change by the method
presented here.  However, in the process of quantization of the
non-abelian gauge theory, while changing from the Hamiltonian
formalism with non-covariant gauge fixing to the Lagrangian
formalism with covariant gauge fixing there are many manipulations
of constraints such as Gauss's law. The treatment of these
constraints is not completely justified from the present point of
view. In many cases the actual, physical, fields are used as
Lagrange multipliers, and a relation such as (\ref{basic}) does not
appear. It is possible that some important piece of information of
the theory is lost in the process.

\section{Discussion}

In this work I developed a formalism in order to treat interactions
between quantum and classical fields through the path integral. It
seems that one is able to describe quantum-classical interactions in
field theory self-consistently with this method. The path integral
formalism is particularly simple and can hopefully be generalized in
other cases involving such interactions.

Some further applications of this work, besides the conceptual
problem of quantum-classical interactions, would be in cases of a
non-renormalizable classical interaction, such as gravity. Note
that, in the example described in Sec.~2,
 there is nothing that demands the classical
field, $\phi$, to have a renormalizable quantum equivalent. Even a
non-renormalizable self-interaction of the classical field does not
generate higher order terms since there are no classical loops.
There may be of course quantum-classical interaction terms that
generate higher order terms, so the problem of renormalizability
then depends on the specifics of the model. Even so, this method may
be useful in treating the stress-tensor renormalization equations
\begin{equation}
G_{\mu\nu}=-8\pi G <T_{\mu\nu}>
\end{equation}
in a self-consistent way.

 I have also given an
example of an application of this work in the case of the
non-abelian gauge field theory by using this method in order to
treat Gauss's law as a classical equation. This is different in
spirit than the previous discussion, it shows, however, another
application of this method in this case, where it can be useful for
studying the infrared properties of the theory. The treatment of the
Coulomb interaction as purely classical in nature has resulted in
the generation of purely quantum terms of the Coleman-Weinberg type
that are related to the confinement mechanism.

Even though the treatment of the non-abelian gauge theory presented
here has a different motivation than previous works, there are
several common features with various other approaches to confinement
\cite{kogut, bander, thooft, zwanziger, c1, c2, c3, c4}. Namely we
find: the appearance of a negative metric propagator (the
$\lambda-\lambda$ line which is crucial in describing the classical
nature); the emergence of a "dynamical" mass term (\ref{res4}) and
an effective potential generated through radiative corrections; an
effective action reminiscent of older phenomenological actions that
were used to describe confinement (although with important
differences described before).

This method can probably be useful in various other problems in
theories that have infrared singularities: one can, presumably, use
this method in order to study the infrared modes of these theories
classically, and include the quantum corrections from higher
momentum modes self-consistently. The problem here, of course, is
the use of an arbitrary cut-off scale, and care should be taken in
order to derive cut-off independent results.

\vspace{0.5in}

\centerline{\bf Acknowledgements} \noindent This work was done in
the National Technical University of Athens. I am grateful to the
people of the Physics Department for their hospitality.

\newpage
\epsffile{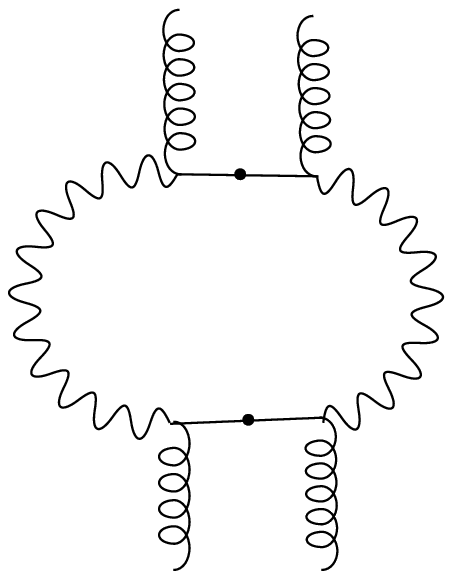}
\bigskip
\bigskip

\noindent Fig.~1: An example of a diagram used for the calculation
of $U(\lambda)$ and the derivation of Eqs.~(\ref{res3}, \ref{res4}).
The wiggly lines denote $A_i$, $A_j$ fields. The curly lines denote
the $\lambda$ field and the solid lines denote $A_0$. The solid
lines with a dot denote the modified $A_0-A_0$ propagator,
$\tilde{G}_{00}$, derived in the text, that does not contain the
Coulomb interaction; the $\lambda-A_0$ and $\lambda-\lambda$ lines
that carry the Coulomb propagator cannot be also added in its place,
because the vertices are only linear in $\lambda$.

\end{document}